\documentclass[aps,pre,amsmath,amsfonts,superscriptaddress,floatfix]{revtex4}

\usepackage{amsmath,amsfonts,amsthm,amssymb}
\usepackage{graphicx}
\usepackage{hhline}
\usepackage{color}
\usepackage{tikz}
\usepackage{soul,xcolor}
\usepackage{enumerate} 
\usetikzlibrary{calc,decorations.markings}

\graphicspath{ {../Images/} }

\def\al{\alpha}

\def\D{\Delta}

\def\p{\phi}
\newcommand*{\pd}[3][]{\ensuremath{\frac{\partial^{#1} #2}{\partial #3}}}
\def\dd{\mathrm{d}}

\begin{document}
	\setstcolor{red}
	
	\title{Quantifying uncertainty in a predictive model for popularity dynamics}
	\author{Joseph D.~\surname{O'Brien}}
	\affiliation{MACSI, Department of Mathematics and Statistics,
		University of Limerick, Limerick V94 T9PX, Ireland}
	\author{Alberto Aleta}
	\affiliation{Institute for Biocomputation and Physics of Complex Systems, University of Zaragoza, Zaragoza 50018, Spain}
	\affiliation{ISI Foundation, 10126 Turin, Italy}
	\author{Yamir Moreno}
	\affiliation{Institute for Biocomputation and Physics of Complex Systems, University of Zaragoza, Zaragoza 50018, Spain}
	\affiliation{ISI Foundation, 10126 Turin, Italy}
	\affiliation{Department of Theoretical Physics, Faculty of Sciences, University of Zaragoza, Zaragoza 50009, Spain}
	\author{James P. Gleeson}
	\affiliation{MACSI, Department of Mathematics and Statistics,
		University of Limerick, Limerick V94 T9PX, Ireland}
	\date{\today}
	
	
	\begin{abstract}
		
		The Hawkes process has garnered attention in recent years for its suitability to describe the behavior of online information cascades. Here, we present a fully tractable approach to analytically describe the distribution of the number of events in a Hawkes process, which, in contrast to purely empirical studies or simulation-based models, enables the effect of process parameters on cascade dynamics to be analyzed. We show that the presented theory also allows predictions regarding the future distribution of events after a given number of events have been observed during a time window. Our results are derived through a differential-equation approach to attain the governing equations of a general branching process. We confirm our theoretical findings through extensive simulations of such processes. This work provides the basis for more complete analyses of the self-exciting processes that govern the spreading of information through many communication platforms, including the potential to predict cascade dynamics within confidence limits.
		
	\end{abstract}
	
	\maketitle
	
	\section[Intro]{Introduction}\label{sec1}
	
	The ease with which individuals may now access online content has revolutionized the way in which information is consumed \cite{lorenz2019accelerating} with social media platforms and online opinion boards \cite{matsa2018news, kwak2010twitter, medvedev2017anatomy, aragon2017thread} being two paradigmatic examples of how information is generated, transmitted and finally absorbed. Modern information communication allow users to both become informed on topics and easily interact online with each other, which may lead to further discussion in a cascade like manner. This results in the occurrence of tree-shaped cascades, the size of which describes the popularity of a discussion. This size metric is one of the key elements in the modern-day social media sphere and can range from a single post in common discussions to values covering several orders of magnitude for viral events, in an extremely heterogeneous fashion \cite{vosoughi2018spread, goel2016structural, lerman2010information}.
	
	One aspect of modeling information cascades that is of paramount importance, not only on opinion boards but on any type of online media, is the predictability of how content popularity changes over time \cite{Cheng_cascades14, miotto2014predictability,Weng_Virality2013,bandari2012pulse,szabo2010predicting, mishra2016feature}. The ability to accurately forecast which online content will become popular has attracted a great deal of attention for several reasons. First, the sheer volume of content that is now available through online platforms, all of which are competing for the limited attention of the users \cite{GleesonPRL14,gleeson2016effects,obrien2019spreading,weng2012competition}, has created a demand for techniques that can decide the order in which users observe new posts, something that is affected by the popularity of the information and at the same time affects that same popularity. This is sometimes referred to as algorithmic bias, by virtue of which more popular topics get more exposure and their popularity is further reinforced. Second, the capacity to predetermine whether or not content will go viral would be an extremely advantageous proposition for corporations, who spend enormous amounts of capital on such campaigns in the hope of increasing their market share \cite{morrissey2007clients, van2010viral, van2010viral_a}. 
	
	Recent literature has focused on developing methods based upon the theory of self-exciting point processes \cite{daley2003introduction}. Unlike homogeneous Poisson processes, the occurrence of previous events in such processes increases the future rate of activity, which can produce a ``snowball effect,'' leading to heavy-tailed distributions of cascade sizes, as observed in empirical data \cite{lerman2012social,dow2013anatomy,Medvedev_reddit18}. An example of such a process is the Hawkes process \cite{hawkes1971spectra}, the application of which has been used to both model and predict the evolution of retweet cascades on Twitter by incorporating the underlying network topology along with machine learning techniques \cite{zhao2015seismic,kobayashi2016tideh}, the popularity of threads on Reddit \cite{Medvedev_reddit18, krohn2019modelling} and predicting the number of views Youtube videos receive as a result of the discussion taking place on other social media platforms \cite{rizoiu2017expecting}. While these works all highlight the capability of the Hawkes process in predicting popularity cascades of online content, there is a lack of knowledge regarding how the predictability of these cascades is affected by the underlying parameters of the Hawkes process.
	
	The aim of this article is to first introduce an analytically tractable approach to fully describe the Hawkes process using the theory of
	branching processes, which has been successfully applied to other spreading processes on online media platforms \cite{GleesonPRL14,gleeson2016effects,IribarrenPRL09,Iribarren2011_PRE,obrien2019spreading}. We obtain the expressions governing the process through an alternative differential-equation framework. Using the results obtained from this analysis, we then turn our attention to the question of predictability in systems with dynamics of this type. We show that our probabilistic interpretation enables the calculation  not only of analytical predictions of future popularity but also of levels of confidence in these values directly from the model itself.  It should be noted that while we interpret the quantity of interest in this work to be the popularity of online content, our general analytic derivation is not restricted to these types of systems. More classically, Hawkes processes have been used to describe a wide range of systems, such as those found in finance \cite{bowsher2007modelling, bacry2015hawkes, bacry2014hawkes, embrechts2011multivariate, rambaldi2015modeling, ait2015modeling}, neuronal activity \cite{pernice2011structure,gerhard2017stability}, or seismology \cite{ogata1988statistical, ogata1998space, saichev2006universal, shcherbakov2019forecasting}. Hence, the expressions obtained in the following can be applied to any process described by such dynamics simply by changing the phrase discussion tree for point process and post for event. 
	
	Before proceeding, we briefly review some of the existing literature that has also considered the use of branching processes in describing such models. The description of an opinion board, specifically Reddit, and also the question of predictability were considered in Ref.~\cite{Medvedev_reddit18}, wherein parameters estimated from the thread's history were used to calculate the expected number of events in future thread activity. 
	Multiple works have also incorporated other elements of online platforms into their models to assist in predicting cascade sizes. For example, Refs.~\cite{mishra2016feature,rizoiu2017hawkes} describe Twitter cascades, via extensive numerical predictions, using the follower network of the tweeting individual. In Ref.~\cite{saichev2013fertility} the authors consider a specific Hawkes process to obtain, through a branching process description, the distribution of events in a certain time window and the distribution of recurrence times between events, without commenting on the predictability of such a model after some observation period. 
	
	Our work differs from the aforementioned literature in several ways. First, we consider the most generic form of Hawkes process (unlike, for example, \cite{rizoiu2017hawkes} which makes use of a given memory kernel, or \cite{saichev2013fertility}, which considers the constant background intensity case) to obtain equations that fully describe the distribution of events at any time; the generality of this approach lends itself to application in a wider number of areas. Second, when we turn to the question of predictability of such a process, which was not considered in \cite{saichev2013fertility}, we introduce a mathematically tractable model which allows one to calculate the expected size of the underlying tree for any given process while also opening up the possibility of determining prediction intervals for the number of events to occur by a given time. This model thus allows the user to provide a level of confidence in their predictions directly from theory rather than via simulation of an extensive number of ensembles. 

	The remainder of this paper is organized as follows. In Sec.~\ref{sec:model}, we introduce a branching process description of the Hawkes process derived through a differential equation formulation, which is then used to determine properties of cascades generated by such processes. Specifically, our approach allows us to fully determine the distribution of the number of events in a given process at any point in time. Having described the general Hawkes process, in Sec.~\ref{sec:prediction} we consider the predictability of such processes after observing a cascade over a certain time period. We focus on the case whereby we have observed a given process for some period of time, which we call the observation window and then, based on the occurrence of events within this window, our theory enables the prediction of not only the expected number of events by some future time, but also the entire distribution of events at each point in the future. In Sec.~\ref{sec:simulations} we perform extensive numerical simulations with the aim of validating our model's prediction. We present a summary and our conclusions in Sec.~\ref{sec:conclusions}.
	
	\section[Model]{Hawkes Process Model}\label{sec:model}
\begin{figure}[ht!]
		\centering
		\includegraphics[width = 0.75\textwidth]{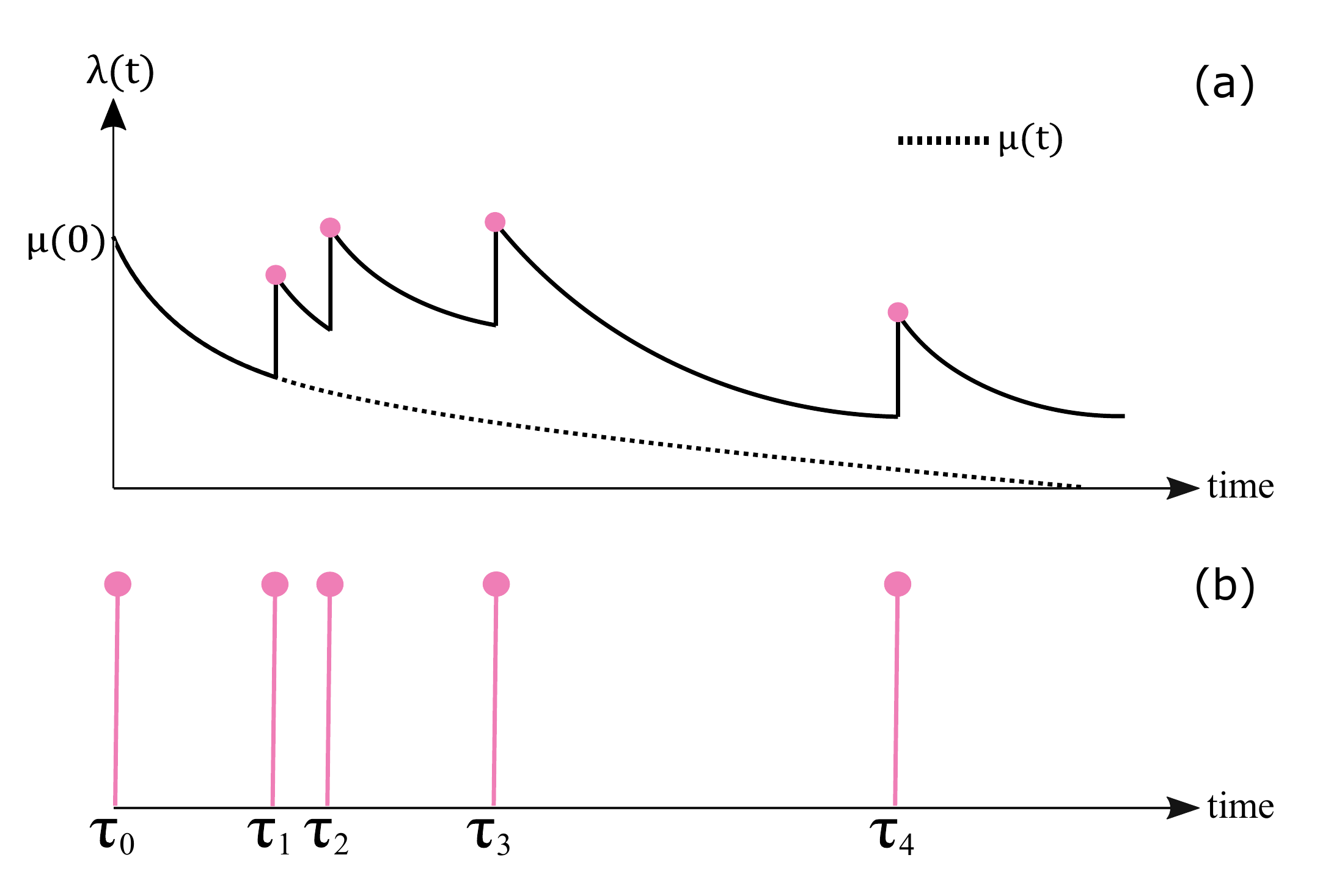}
		\caption{Example of the Hawkes process. (a) Intensity function $\lambda(t)$ given by Eq.~\eqref{hawkes_eq} along with the background intensity $\mu(t)$ without any self-excitation, shown by the dashed line. (b) Corresponding event sequence of five events at times $\tau_i$.}
		\label{fig:hawkes_desc}
	\end{figure}

	Self-exciting processes are those in which the likelihood of a stochastic event taking place increases with the occurrence of past events. Hence, the probability of an event occurring in these processes is determined by a time-dependent intensity, $\lambda(t)$, defined such that $\lambda(t) \, dt$ is the expected number of events in the interval $[t, t + dt]$.~One particular example of these processes is the Hawkes process \cite{hawkes1971spectra}. In this class of point processes the intensity is given by
	\begin{equation}
	\lambda(t) = \mu(t) + \xi \sum_{\tau_i < t}\p(t-\tau_i),
	\label{hawkes_eq}
	\end{equation}
	where $\mu(t)$ is known as the background intensity, describing the likelihood of an exogenous event occurring independently of other events. The non-negative $\phi(t)$ term is known as the memory kernel or the excitation function, as it gives the amount by which previous events that have occurred at times $\tau_i < t$ increase the probability of occurrence of future events. As such, it accounts for the endogenous or self-exciting part of the process. Further, we normalize $\phi(t)$ such that $\int_0^\infty \phi(t) dt = 1$. Note that with this interpretation, one may consider $\phi(t)$ as a probability density function or a memory-time distribution \cite{gleeson2016effects}. Finally, the parameter $\xi$ may be thought of as the branching number of the process, whereby it defines the average number of events caused as a result of an event and may in some sense capture the fitness of a previous event. Note that in the case $\xi = 0$, one recovers an inhomogeneous Poisson process. In this article we specifically focus on the subcritical case where $\xi < 1$, which results in finite-size (albeit potentially very large) cascades. In the alternative supercritical case ($\xi > 1$) the average number of events grows exponentially without bound, which is evidently meaningless for the processes of interest in this work. Figure \ref{fig:hawkes_desc} shows an example realization of a Hawkes process.
	
	To describe this process further we will proceed to consider the behavior of a discussion tree on a social media platform. Suppose that a discussion is started by a user posting at time $\tau_0 = 0$. This comment will be referred to as the seed of the discussion. The size of the discussion can increase in two ways, as described in \cite{Medvedev_reddit18}. First, it may receive new comments itself, which are essentially replies to the seed, with rate $\mu(t)$. This rate can depend on a number of factors, such as the discussion's position on the ``front page'' of the site \cite{hogg2012social, wu2007novelty} or the number of users present on the site in the period after the tree was created. Second, comments made to the seed can themselves receive replies and thus be viewed as subtrees of the tree that represents the whole discussion. A comment made at time $\tau_i$ receives replies with intensity $\xi \, \phi(t - \tau_i)$, where $\xi$ can be viewed as some fitness parameter which could depend on factors such as the quality of the content in the comment, the time of day at which it first appeared, or the popularity of the individual who made the comment. For simplicity, however, we will consider it as a parameter unique to each discussion.
	
\begin{figure}[ht!]
		\centering
		\includegraphics[width = 0.75\textwidth]{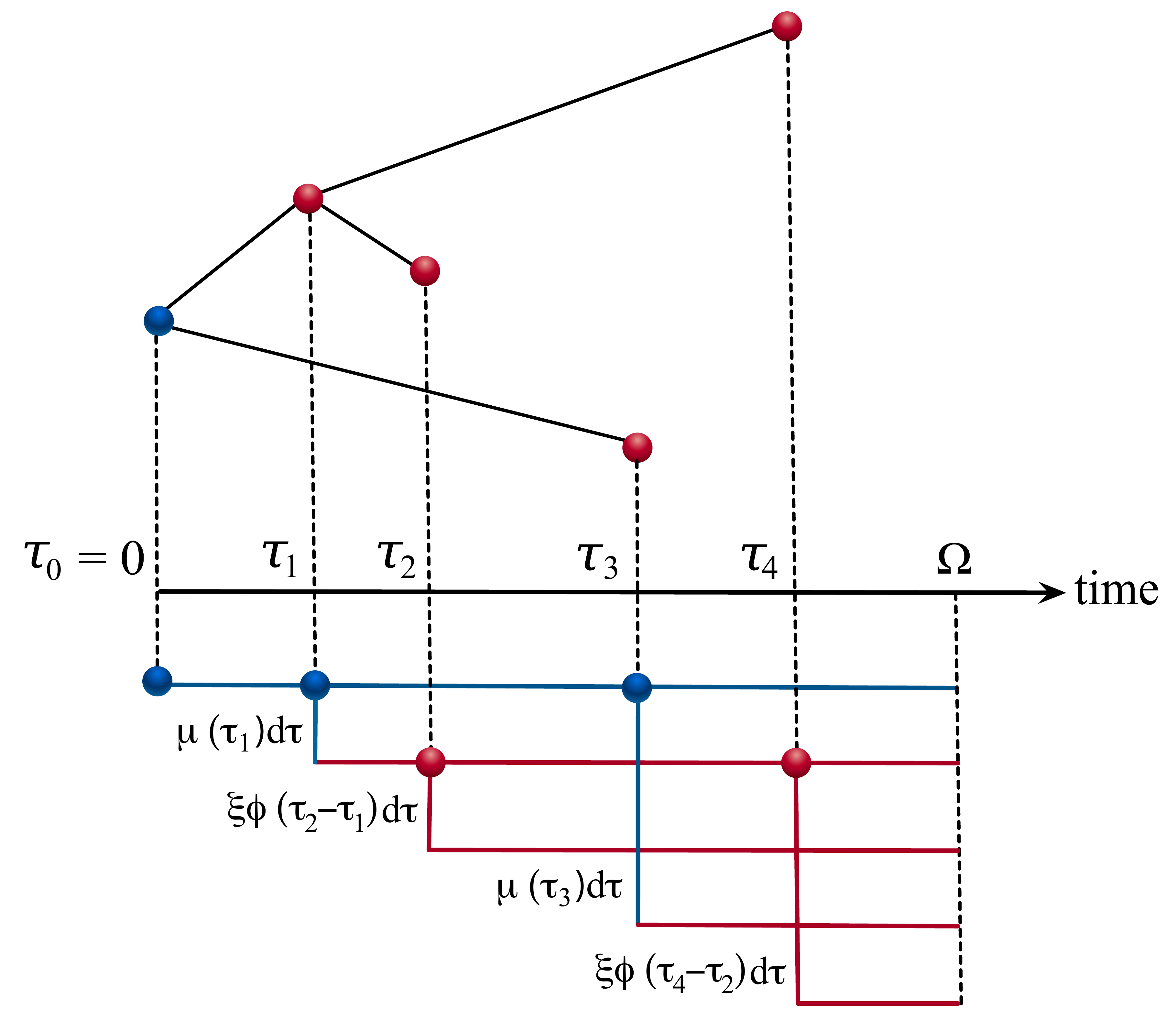}
		\caption{Schematic of the model. A discussion is started by the seed at time $\tau_0 = 0$, represented by the horizontal blue line. At time $\tau_1$ the seed receives a reply, shown by the blue dot, with probability $\mu(\tau_1) \, d\tau$ which starts its own subtree shown by the first horizontal red line. At time $\tau_2$ the comment receives a reply highlighted by the red dot, with probability $\xi \, \p(\tau_2 - \tau_1) \, d\tau$. The seed receives another reply at time $\tau_3$, w.p. $\mu(\tau_3) d\tau$, which again starts another subtree. Finally, the subtree started at time $\tau_1$ receives another reply at time $\tau_4$, which occurred with probability. $\xi \, \p(\tau_4 - \tau_1) \, d\tau$, itself starting a new branch. We now observe the entire tree size, i.e., the number of comments, to be $5$ at time $\Omega$.}
		\label{hawkes_time}
	\end{figure}

	We can graphically arrange the posts in a discussion by the shape of a tree in which the nodes represent posts and an edge is established between two nodes if one is a reply to the other (see Fig.~\ref{hawkes_time}). This representation suggests that the dynamics of such a system can be described as a branching process. Letting $q_m(t)$ be the probability that a discussion of age $t$ has popularity $m$, i.e., it has received $m$ comments, we may define the probability generating function (PGF) \cite{wilf2005generatingfunctionology} of the popularity distribution as
	\begin{equation}
	H(t;x) = \sum_{m = 1}^{\infty}q_m(t)\, x^m \,.
	\label{gen_fun}
	\end{equation}
	In passing we mention one important property of PGFs that we will frequently use in the following, namely, that the PGF for the sum of two random variables $X$ and $Y$ is simply the product of their individual generating functions.
	For the remainder of this section, we will focus on a derivation of this PGF through a branching process description of the Hawkes process \cite{Athreya2004, Harris2002} followed by an analysis of some of the properties that can be extracted from this interpretation.
	
	\subsection{Branching Process Description of the Hawkes Process}\label{sec:BPapprox}
	
	We propose a derivation of the equations for the branching process generating function given by Eq.~(\ref{gen_fun}) that is based on considering the change in the PGFs over infinitesimal time intervals. This leads naturally to a differential equation formulation. In the Appendix, we show how the method introduced below can also be used to derive the well-known integral equation for the Bellman-Harris process \cite{Athreya2004,Harris2002,IribarrenPRL09,Iribarren2011_PRE}.
		
	Let us first consider the distribution of the size of subtrees described above. Suppose that a new comment arrives at the subtree at time $\tau$. We define the PGF of the resulting size distribution as $G(\tau,a,\Omega;x)$, where $a$ is the age of the comment that originated the subtree and $\Omega$ the time at which the distribution is observed. Note that the time of origin of the subtree is thus $c \equiv \tau - a$. Now suppose we examine how this PGF changes over a time interval $(\tau-\D t, \tau)$. If $\D t$ is sufficiently small so that at most one event occurs in a time interval of length $\Delta t$, there are two possible events that may occur.
	\begin{enumerate}[(i)]
		\item A user may decide to reply to the comment of age $a$ with probability $\xi \, \p(a) \D t$, which depends on both the time elapsed since the comment was made and its fitness. This will result in a new comment of age $0$ which may, in turn, receive its own replies. Concurrently, the original comment may also spawn further replies, so the total contribution to $G(\tau - \D t, a - \D t, \Omega; x)$ is $G(\tau,0,\Omega;x)G(\tau,a,\Omega;x)$, where we have made use of the aforementioned property that the PGF for the sum of two random variables is simply the product of their respective PGFs.
		\item There may also be no replies to the comment in this time interval. This occurs with probability $1 - \xi \, \p(a) \D t$, and the contribution to $G(\tau - \D t, a - \D t, \Omega; x)$ in this case is simply $G(\tau,a,\Omega;x)$.
	\end{enumerate}
	Thus, the equation governing the change in $G$ over the interval  $(\tau - \D t, \tau)$ is given by
	\begin{equation}
	G(\tau - \D t, a - \D t, \Omega; x) = \xi \, \p(a) \D t \, G(\tau,0,\Omega;x)G(\tau,a,\Omega;x) + \left[1 - \xi \, \p(a) \D t\right]G(\tau,a,\Omega;x).
	\end{equation}
	This equation may then be expressed as a partial differential equation by considering the two dimensional Taylor approximation in the small-$\D t$ limit to obtain
	\begin{equation}
	\pd{G}{\tau} + \pd{G}{a} = \xi \, \p(a)\left[1 - G(\tau,0,\Omega;x)\right]G,
	\end{equation}
	where we write $G$ for $G(\tau,a,\Omega;x)$. This equation may be simplified via the method of characteristics \cite{haberman1983elementary} (letting $c = \tau - a$ be constant along a characteristic) to obtain the ordinary differential equation
	\begin{equation}
	\frac{dG}{d\tau} = \xi \, \p(\tau - c) \left[1 - G(\tau,0,\Omega;x)\right] G,
	\end{equation}
	for the change in $G$ along a characteristic, which we may solve by noting that $G(\Omega, \Omega - c, \Omega; x) = x$, i.e., the initial subtree size when a comment is made is 1, to obtain
	\begin{equation}
	G(\tau,\tau - c,\Omega;x) = x \, \exp\left\{-\xi \int_{\tau}^{\Omega}\p(v - c)\left[1 - G(v,0,\Omega;x)\right] \, dv \right\}.
	\label{Gintegral}
	\end{equation}
	We may now insert $a = \tau - c$, and introduce the ``tree age'' $t = \Omega - \tau$, i.e., the age of the subtree at observation time, and let $w = \Omega - v$, which gives
	\begin{equation}
	G(t,a;x) = x \, \exp\left\{\xi\int_{0}^{t}\p(w + a)\left[G(t-w, 0; x) - 1\right] \, dw \right\}.
	\end{equation}
	Finally, we let $a = 0$ to obtain the PGF for the subtree size at age $t$ as a result of a new post (and dropping the $a=0$ argument from $G$)
	\begin{equation}
	G(t;x) = x \, \exp\left\{\xi \int_{0}^{t}\p(w)\left[G(t-w; x) - 1\right] \, d w \right\}.
	\label{eqG}
	\end{equation}
	
	We will now proceed to determine the PGF of the entire size distribution of a tree with age $t$, $H(t;x)$. Equation \eqref{eqG} describes the distribution of the size of a subtree as a result of replying to either the original seed or another comment (which would be itself a root of a previous subtree). To determine the entire tree size distribution we first define $H(\tau,\Omega;x)$ to be the PGF of a cascade that had started at time $0$, observed over the time interval $(\tau,\Omega)$. As above, we consider the two possible events which may occur in relation to the seed over the small time interval $(\tau - \D t, \tau)$.
	\begin{enumerate}[(i)]
		\item There may be a comment made on the post (i.e., in direct reply to the seed) with probability $\mu(t) \D t$, resulting in a subtree that develops from this comment with size distribution determined by $G(\tau,0,\Omega;x)$. The seed itself may also receive further comments. The total contribution to $H(\tau-\D t,\Omega;x)$ is given by the product of the two individual PGFs $G(\tau,0,\Omega;x)H(\tau,\Omega;x)$.
		\item The seed may receive no comments in this time interval which occurs with probability $1 - \mu(\tau) \D t$, so the contribution to $H(\tau-\D t,\Omega;x)$ is simply $H(\tau,\Omega;x)$.
	\end{enumerate}
	Analogously to the previous PGF, we take the $\D t \rightarrow 0$ limit and use the method of characteristics to obtain the differential equation
	\begin{equation}
	\frac{d H(\tau,\Omega;x)}{d\tau} = \mu(\tau) \, \left[1 - G(\tau,0,\Omega;x)\right] \, H(\tau,\Omega;x),
	\end{equation}
	noting that the initial tree size is 1 (the seed itself), i.e., $H(0,0;x) = x$. We remark that $\Omega$ is in fact the age of the post. For consistency, we will define this to be $t$ and set $\tau = 0$ such that we observe the entire tree, obtaining
	\begin{equation}
	H(t;x) = x \, \exp\left\{\int_{0}^{t} \mu(y)\,\left[G(t-y;x) - 1\right] \, d y \right\},
	\label{eqH}
	\end{equation}
	where $G(t-y;x)$ is given by Eq.~\eqref{eqG}. Therefore, the entire behavior of the Hawkes process is described by Eqs.~\eqref{eqG} and \eqref{eqH}. Note that a similar expression was obtained by Hawkes in the case of constant background intensity, i.e., $\mu(t) = \lambda_0$, by considering the process as a Poisson cluster process \cite{hawkes1974cluster}; however, a description of the most general case including time-varying background intensity is lacking.
	
	
	\subsection[Analysis]{Mathematical analysis and properties of the branching tree}\label{sec3}
	
	\subsubsection{Mean Size of a Cascade}\label{sec:mean}
	
	To determine the expected size of a discussion with age $t$, $m(t)$, we use the property that
	\begin{equation}
	m(t) = \sum_{n = 1}^{\infty} n \, q_n(t) = \left.\pd{H(t;x)}{x}\right|_{x=1}\,.
	\end{equation}
	We then differentiate Eq.~\eqref{eqH} and evaluate at $x = 1$ to obtain an integral equation for $m(t)$,
	\begin{equation}
	m(t) = 1 + \int_{0}^{t} \mu(y) \, m_G(t-y) \, d y \,,
	\label{mean_h}
	\end{equation}
	where $m_G(t)$ represents the mean size of a subtree with age $t$ and is defined by $m_G = \left.\frac{\partial G}{\partial x}\right|_{x=1}$. Analogously, differentiating Eq.~\eqref{eqG} yields
	\begin{equation}
	m_G(t) = 1 + \xi\int_{0}^{t}\phi(w) \, m_G(t-w) \, d w\,.
	\label{mean_g}
	\end{equation}
	The next step is to take the Laplace transform of Eq.~\eqref{mean_h} to obtain
	\begin{equation}
	\hat{m}(s) = \frac{1}{s} + \hat{\mu}(s) \, \hat{m}_G(s),
	\label{Lap_m}
	\end{equation}
	with $\hat{m}_G(s)$ given by the Laplace transform of Eq.~(\ref{mean_g}), which may be solved exactly to obtain
	\begin{equation}
	\hat{m}_G(s) = \frac{1}{s\left[1 - \xi \,\hat{\phi}(s)\right]};
	\end{equation}
	substituting this value in Eq.~\eqref{Lap_m} gives
	\begin{equation}
	\hat{m}(s) = \frac{1}{s} + \frac{\hat{\mu}(s)}{s\left[1 - \xi \, \hat{\phi}(s)\right]}.
	\label{ms}
	\end{equation}
	
	We may determine the limiting value of $m(t)$ as $t \rightarrow \infty$, if it exists, by making use of the final limit theorem \cite{ogata2002modern}
	\begin{align}
	m(\infty) = \lim_{t \to \infty}m(t) =&\; \lim_{s\to 0} s \, \hat{m}(s) \\
	=&\; 1 + \frac{\int_{0}^{\infty}  \mu(t) \,  d t}{1 - \xi}.
	\label{m_large_t}
	\end{align}
	Furthermore, it is also interesting to understand how this limit is approached in the large-$t$ limit. Following \cite{Iribarren2011_PRE, gleeson2016effects}, the calculation depends on the existence (or not) of the value $\al$ that satisfies
	\begin{equation}
	\xi \int_{0}^{\infty} e^{-\al t} \, \phi(t) \, d t = 1.
	\label{malthusian}
	\end{equation}
	This special value of $\alpha$ is known as the Malthusian parameter \cite{Athreya2004}. If this parameter exists then the limit $m(\infty)$ is approached exponentially:
	\begin{equation}
	m(t) \sim m(\infty) + \frac{\int_{0}^{\infty} e^{-\al p} \, \mu(p) \, d p}{\alpha \, \xi \, \int_{0}^{\infty} p \, \phi(p) \, e^{-\al p} \, d p} \, e^{\al t}, \qquad \text{as} \quad t \rightarrow \infty.
	\end{equation}
	However, if there is no solution $\al$ of Eq.~(\ref{malthusian}), which is the case when $\phi(t)$ is a subexponential distribution (e.g., Gamma distribution), then the large-$t$ behavior of $m(t)$ is given by
	\begin{equation}
	m(t) \sim m(\infty) - \frac{\xi \, \int_{0}^{\infty}\mu(p) \, d p}{(1 - \xi)^2}\left[1 - C(t)\right],
	\label{mean_asy}
	\end{equation}
	where $C(t) = \int_{0}^{t}\phi(t) \, dt$ is the cumulative distribution of the memory kernel.
	
	While the above analysis describes the mean behavior of the long-time dynamics, for some choices of the background intensity and memory kernel it is also possible to exactly determine the entire temporal behavior of the mean cascade size. In the following we will explore two of those choices: constant background intensity with exponential memory kernel and constant background intensity with power-law memory kernel.
	
	\noindent\paragraph{Exponential memory kernel with constant background intensity}\hfill \break
	To consider a specific example, we take the well-studied case where the process's memory kernel is exponentially distributed \cite{hawkes1971spectra, hawkes1974cluster, lewis2012self, masuda2013self, aoki2016input}  with mean time $1/\beta$ and background intensity given by a constant, i.e., $\mu(t) = \lambda_0$, so that the intensity is given by
	\begin{equation}
	\lambda(t) = \lambda_0 +  \xi \sum_{\tau_i < t}\beta e^{-\beta(t - \tau_i)};
	\label{expHawkes}
	\end{equation}
	in this case \eqref{ms} becomes
	\begin{equation}
	\hat{m}(s) = \frac{1}{s} + \frac{\lambda_0}{s^2}\left[\frac{\beta + s}{s + \beta(1 - \xi)}\right],
	\end{equation}
	which results in the following solution for the mean cascade size as a function of time:
	\begin{equation}
	m(t) = 1 + \frac{\lambda_0}{1 - \xi}\left[t + \frac{\xi}{\beta(1-\xi)}\left(e^{-\beta(1-\xi)t} - 1\right)\right].
	\label{mean_exp_classic}
	\end{equation}
	Note that we initially have a linear term in $t$, as one would expect in the case of constant background intensity, which gives some constant probability of replies to a post regardless of the time. This growth does eventually slow down as a result of the exponential decay term, but is still determined as linear growth. A detailed analysis into this form of the Hawkes process was considered in \cite{oakes1975markovian}, where it was shown that this simplistic setting is equivalent to a continuous-time Markov process, as one would anticipate from the exponential kernel.	\linebreak	
	\paragraph{Shifted power-law memory kernel with constant background intensity}\hfill \break
	Another frequently used kernel is that of the shifted power-law memory kernel $\phi(t) \propto (t + c)^{-(1+\beta)}$ for $\beta > 0$ \cite{ogata1988statistical,sornette2004endogenous, crane2008robust, jo2015correlated,mishra2016feature, rizoiu2017expecting}. From this kernel, along with a constant background intensity and the normalization condition [i.e., $\int \phi(t) \, d t = 1$] we have
	\begin{equation}
	\lambda(t) = \lambda_0 + \xi \sum_{\tau_i < t} \beta \, c^{\beta} \, (t - \tau_i +c)^{-(1+\beta)} \,.
	\label{PLHawkes}
	\end{equation}
	Hence, in this case $\hat{\mu}(s) = \frac{\lambda_0}{s}$ and $\hat{\phi}(s) = \beta\,c^\beta e^{cs} s^\beta \, \Gamma(-\beta,cs)$, where $\Gamma$ is the upper incomplete Gamma function. We may determine the early-time behavior in this instance by considering $s \gg 1$ and expanding the incomplete $\Gamma$ function giving
	\begin{equation}
	\hat{\phi}(s) \sim \frac{\beta}{c}\frac{1}{s} \,.
	\end{equation}
	Inserting this expression into Eq.~\eqref{ms} and applying the inverse Laplace transform, we obtain
	\begin{equation}
	m(t) \approx 1 + \frac{c\lambda_0}{\xi \beta}\left[e^{(\beta/\xi c)t} - 1\right].
	\end{equation}
	The presence of a constant background intensity makes the question regarding long-time behavior less relevant. This is due to the fact that the constant (non zero) probability of an event to occur results in infinitely large cascades in the large-$t$ asymptotic limit described, on average, by Eq.~\eqref{m_large_t}. If this factor were not constant but rather a time-decaying $\mu(t)$, as the theory allows, the large-$t$ behavior would be governed by Eq.~\eqref{mean_asy}.
	
	\subsubsection{Probability a Discussion Receives no Comments\label{subsubsec:noComments}}
	Another interesting quantity which may be determined directly from the probability generating function is the probability that a discussion tree, once started by a user, receives no responses by the time it has age $t$, denoted given by $q_1(t)$. This value can be obtained from Eq.~(\ref{eqH}) as follows:
	\begin{align}
	q_1(t) &= \lim\limits_{x \to 0} \frac{H(t;x)}{x} \\
	&=  \lim\limits_{x \to 0} \exp\left\{\int_{0}^{t} \mu(y)\,\left[G(t-y;x) - 1\right] \, d y \right\} \\
	&= \exp\left\{-\int_{0}^{t} \mu(y)\, d y \right\}.
	\label{q1classic}
	\end{align}
	As one would expect, this quantity is purely determined by the background intensity function as the memory kernel only becomes a factor once the seed has received a reply.
	
	{\subsubsection{Distribution of the number of events}\label{sec:IFFT}
		
		The branching process interpretation of a given Hawkes process described in Sec.~\ref{sec:BPapprox} allows one to determine the entire dynamics of the given process through its PGFs \eqref{eqG} and \eqref{eqH}. In order to extract $q_m(t)$ $-$that is, the probability that a tree has size $m$ at age $t$$-$ from these equations, one is required to differentiate the PGF $H(t;x)$ $m$ times, i.e., 
		\begin{equation}
		q_m(t) = \left.\frac{1}{m!}\frac{\partial^m}{\partial x^m} H(t;x) \right|_{x=0};
		\end{equation}
		the numerical differentiation required to perform this calculation is however inaccurate for large-$m$ values and as such we instead use the Cauchy formula \cite{cavers1978fast} for the derivative of a function to obtain
		\begin{equation}
		q_m(t) = \frac{1}{2\pi i}\oint_C H(t;x)x^{-(m+1)} \, d x,   
		\end{equation}
		where $C$ is a contour in the complex-$x$ plane such that all poles of $H(t;x)$ lie outside $C$. This may then be evaluated in a numerically accurate manner through inverse fast Fourier transform routines \cite{GleesonPRL14,cavers1978fast,abate1992fourier} to allow one to obtain the distribution of tree sizes at a given time $t$.
		
		\subsection{Numerical Simulations}\label{num_sim_1}
		
		In Fig.~\ref{fig:sim1} we compare the theoretical predictions with numerical simulations of the Hawkes process. In Figs.~\ref{fig:sim1}(a) and \ref{fig:sim1}(c) we consider the case of constant background intensity and exponential memory kernel. Specifically, in Fig.~\ref{fig:sim1}(a) the expected number of events is shown for a number of different fitness parameters ($\xi = 0.2, 0.5, 0.8$) along with $\lambda_0 = 0.1$ and $\beta = 3$. The lines correspond to the theoretical expected value of Eq.~\eqref{mean_exp_classic}, while the circles represent the average number of events over an ensemble of $10^6$ realizations. As expected, larger fitness values result in a greater expected number of events. Note also that for large values of $t$, a linear behavior of the expected value is recovered, as suggested by the theoretical result of Eq.~\eqref{mean_exp_classic}.
		
		In each case we extract the distribution of tree sizes at each time point via the approach described in Sec.~\ref{sec:IFFT}. Figure~\ref{fig:sim1}(c) shows the corresponding complementary cumulative distribution function (CCDF), which is the probability of having $m$ or more comments on the discussion, following the process described above, at several different ages. Regarding these results, it is clear that the likelihood of a larger number of events is increasing with time.
		
		Figures~\ref{fig:sim1}(b) and ~\ref{fig:sim1}(d) show the equivalent tests for the case of the power-law memory kernel given by Eq.~\eqref{PLHawkes} with $\beta = 0.3$ and $c = 0.01$. Note that although there is no closed-form expression for the expected value in  Fig.~\ref{fig:sim1}(b), we can determine these values directly from the distribution function obtained through the PGF of the process. We find the results to be in strong agreement with simulations (in spite of the limitations of discrete-time simulations of such dynamics \cite{fennell2016limitations}) therefore validating the branching process interpretation.
		
		One may note that the expected size of cascades shown in Fig.~\ref{fig:sim1} is not converging to a fixed value; this is due to the constant background intensity which results in some probability of an event occurring at each point that leads to the cascades growing, on average, without bound as suggested by Eq.~\eqref{m_large_t}. If instead the background intensity is decaying with time, the possibility of convergence on average to a fixed value is more evident. In Fig.~\ref{fig:simSE} we consider a purely self-exciting process such that $\lambda(t) = \xi \phi(t)$, specifically the same exponential memory kernel as in Fig.~\ref{fig:sim1} with $\xi = 0.8$ is used. Figure~\ref{fig:simSE}(a) shows the expected number of events and also the 95\% prediction intervals i.e., the values within which we would expect 95\% of ensembles appeared. The dashed color lines show the equivalent percentiles from simulations. We note that in this scenario these quantities converge at large time; for example, the expected number of events tends towards 5 as Eq.~\eqref{m_large_t} indicates. Figure~\ref{fig:simSE}(b) shows the corresponding CCDFs at each of the time-points considered in Fig.~\ref{fig:simSE}(a). We also comment on the wideness of the prediction intervals describing the process which highlight how they are extremely heterogeneous in comparison to the expected number of events. This highlights the inherent stochasticity underlying processes described by such dynamics, which further emphasizes the importance in obtaining levels of confidence in predictions aside from simply the expected number of events that appears to be an insufficient predictor by itself, which is particularly important when such methods are applied to empirical data.
		
		\begin{figure}[ht!]
			\centering
			\includegraphics[width=\linewidth]{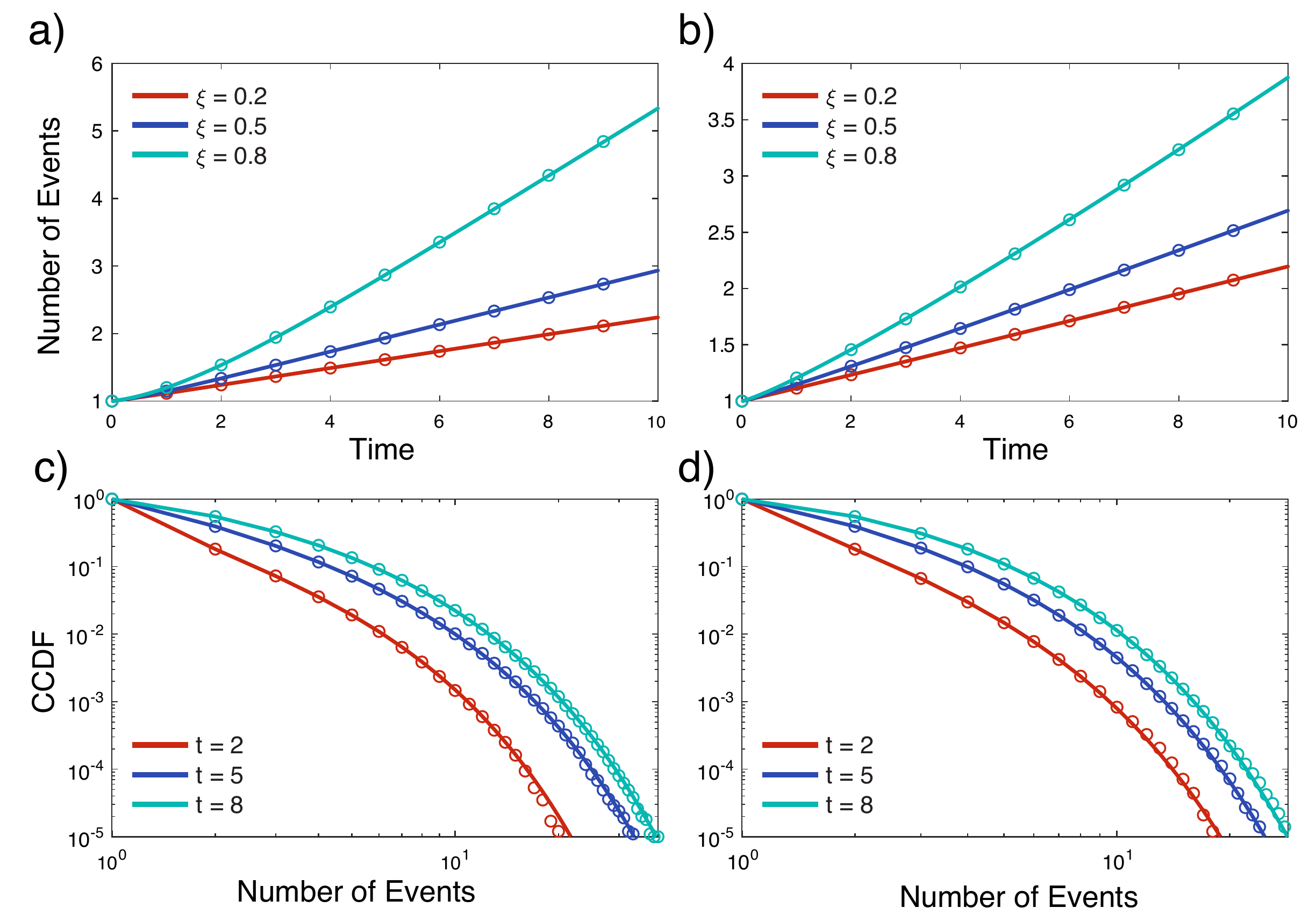}
			\caption{Numerical simulation of the Hawkes process over $10^6$ realizations. (a) Mean number of events for a Hawkes process with exponential memory kernel ($\beta = 3$) and constant background intensity $\lambda_0 = 0.1$ for a range of fitness values. The lines represent the theoretical values for the expected number of events given by Eq.~\eqref{mean_exp_classic}, while the circles represent the corresponding results from simulation. (c) The CCDF of the number of events in a scenario where $\xi = 0.5$ and $\beta = 3$ for three different values of observation time. Again, the lines represent the theoretical CCDF from inversion of the PGF and circles represent the results from simulation. (b) and (d) Corresponding plots for a power-law memory kernel with $c = 0.01$ and $\beta=0.3$.}
			\label{fig:sim1}
		\end{figure}
		
		\begin{figure}[ht!]
			\centering
			\includegraphics[width=\linewidth]{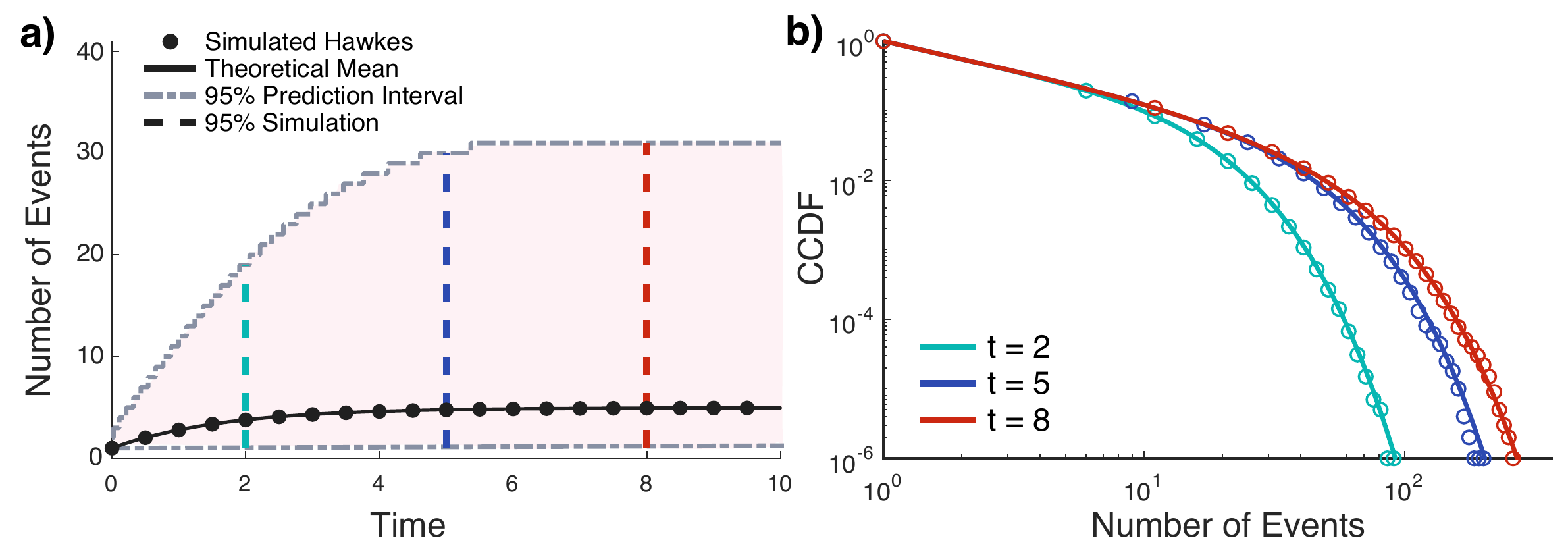}
			\caption{Numerical simulation of the Hawkes process over $10^6$ realizations. (a) Mean number of events for a purely self-exciting Hawkes process with exponential memory kernel ($\beta = 3$) and equivalent background intensity with a fitness value of $\xi = 0.8$. The black solid line represents the theoretical values for the expected number of events while the circles represent the corresponding results from simulation. The gray dash-dotted lines indicates the 95\% prediction interval, i.e., the values between which 95\% of the distribution is centered at each time point, while the colored dashed lines indicate the equivalent percentiles from simulation at three time points $t = 2,5,8$. (b) The CCDF of the number of events at the same three time points.}
			\label{fig:simSE}
		\end{figure}
		
		\section{Predictability of Discussion Popularity}\label{sec:prediction}
		We now focus on the issue of predicting the time-dependent discussion size given that its popularity has been observed from the moment it was first created at time $\tau_0 = 0$, until a time $T$, i.e., over the tree age interval $[0, T]$, which we will refer to as the observation window (see Fig.~\ref{fig:prediction_scheme}). Let us assume that we have observed $n$ comments over this interval at times $\{\tau_0,\tau_1,\dots,\tau_{n-1}\}$ and that we now wish to consider how this process evolves until a new observation at time $\Omega$, i.e., the tree age interval $(T,\Omega]$, which we refer to as the prediction window. Consistent with the branching process equivalence for the Hawkes process, we may treat each of these subtrees seeded during the observation window as independent and we only consider what occurs in that subtree after the time $T$. This means that while a comment may have resulted in a new subtree before time $T$, we treat both of them independently when they develop after time $T$ such that at time $T$ we have $n$ subtrees all of size one but with different ages given by $T - \tau_i$, where $\tau_i$ denotes the time when the subtree was seeded. 
		
		\begin{figure}[ht!]
			\centering
			\includegraphics[width=\linewidth]{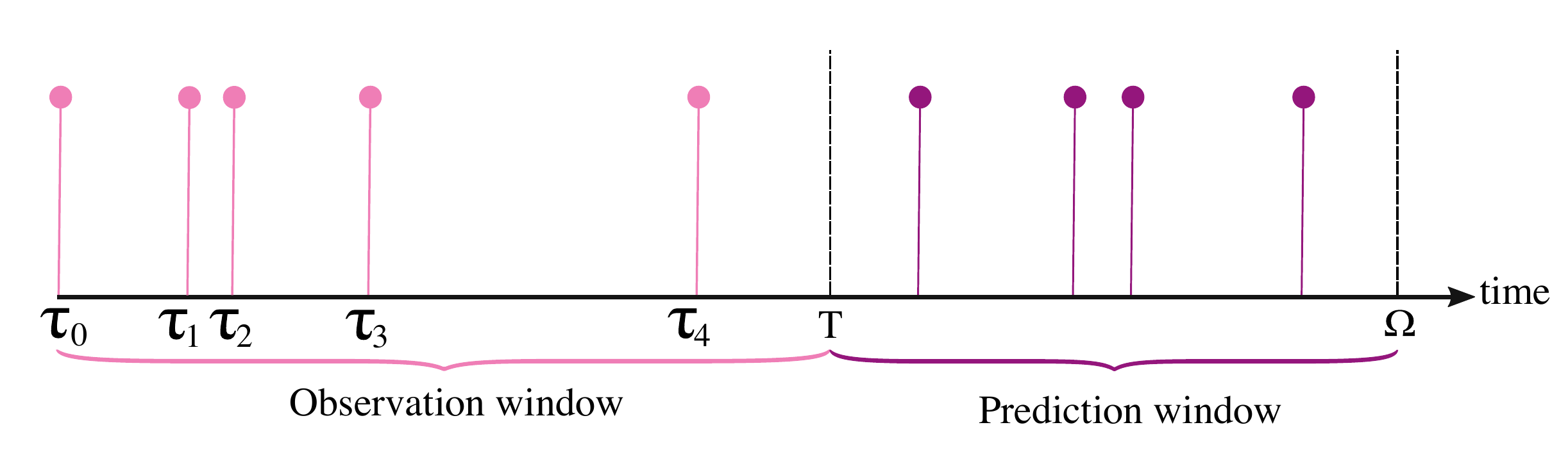}
			\caption{Schematic of the model described in Sec.~\ref{sec:prediction}. For a given process, we observe its evolution over what we refer to as the observation window from the time $\tau_0$ of the seed comment until a future time $T$. During this window we observe $n = 5$ events, including the seed event itself, at times \{$\tau_0, \tau_1, \dots, \tau_4$\}. Our model then considers how the age of each of these events, $a_i = T - \tau_i$, influences the distribution of number of events over a prediction window of length $r = \Omega - T$, where $\Omega$ is the time at which we observe what has occurred in this window. In the ensemble shown above four further events took place by the observation time.}
			\label{fig:prediction_scheme}
		\end{figure}
		
		As in Sec.~\ref{sec:BPapprox}, we have to consider two different types of events. First, we have the seed which started the discussion. Then the remaining $n-1$ comments are replies to the seed or to previous posts on the discussion. We again start by defining $G(\tau,a,\Omega;x)$ to be the PGF of the size distribution of a subtree, seeded as a result of a reply at time $\tau$ to a comment which had age $a$ at this time, observed at time $\Omega$. We are now interested in predicting the dynamics of these $n-1$ subtrees over the time interval $(T,\Omega]$. To do so, let us consider the $(i+1)th$ comment which was made at time $\tau_i < T$ and study the size of the subtree resulting from a reply to this comment at time $\tau$. Therefore, Eq.~\eqref{Gintegral} in this case is 
		\begin{equation}
		G(\tau,a,\Omega;x) = x \, \exp\left\{-\xi \int_{\tau}^{\Omega}\p(v + a - \tau)\left[1 - G(v,0,\Omega;x)\right] \, d v \right\}\,.
		\end{equation}
		The age of the original comment at this time will be $a = \tau - \tau_i$. As we are interested in considering the tree size over the entire prediction window, we set $\tau = T$ so that
		\begin{equation}
		G(T,T - \tau_i,\Omega;x) = x \, \exp\left\{-\xi \int_{T}^{\Omega}\p(v - \tau_i)\left[1 - G(v,0,\Omega;x)\right] \, d v \right\}.
		\end{equation}
		Next, we define the prediction age, i.e., the length of the prediction window, as $r = \Omega - T$. Similarly, letting $w = \Omega - v$, we can rewrite the function in terms of tree age to abridge notation, obtaining
		\begin{equation}
		G(r,T - \tau_i;x) = x \, \exp\left\{-\xi \int_{0}^{r}\p(r+T-\tau_i-w)\left[1 - G(w,0;x)\right] \, d w \right\}.
		\label{Gobs}
		\end{equation}
		We note that the $G$ function within the integral, i.e., for a comment that was made during the prediction window, is exactly of the form described by Eq.~\eqref{eqG}. The $G$ function described by the above equation, however, has an explicit dependence on the age of the comment when we stop observing, $a_i = T - \tau_i$. Thus, in order to compress notation, we define $G_i(r;x)=G(r,a_i;x)$ to highlight the difference between trees started during the observation and prediction windows.
		
		We must also consider replies to the seed comment over this interval. A similar line of thought allows one to see that the PGF describing the distribution of these replies is given by
		\begin{equation}
		H(r;x) = x \, \exp\left\{-\xi \int_{0}^{r}\mu(r+a_0-w)\left[1 - G(w,0;x)\right] \, d w \right\},
		\label{Hobs}
		\end{equation}
		where $a_0$ is simply the age of the discussion when we stop observing, i.e., $a_0 = T$.
		
		Finally, we require one more PGF to completely define the evolution of this process. We denote by $I\left(\{a_0,a_1,\dots,a_{n-1}\}, r; x\right)$ the PGF of total discussion tree size after a prediction window of length $r$, where we have observed $n$ comments on the tree which had ages $\{a_0,a_1,\dots,a_{n-1}\}$ when we stopped observing (i.e., at $t=T$). Therefore, it can be expressed as
		\begin{align}
		I\left(\{a_0,a_1,\dots,a_{n-1}\}, r; x\right) &= H(r;x)\prod_{i = 1}^{n-1}G_i(r;x), \\
		&= x^n \, \exp\left[-\int_{0}^{r} \left(\mu(r+a_0-w) + \sum_{i=1}^{n-1}\, \xi \,\p(r+a_i-w)\right)
		\left[1 - G(w,0;x)\right] \, d w\right],
		\label{Ipgf}
		\end{align}
		where the product of generating functions arises as we are summing the contributions from each of the subtrees started before time $T$. Note that this equation reduces to Eq.~(\ref{eqH}) in the case where we observe only one event ($n = 1$), which has age $0$ when we stop observing ($a_i = 0$) and letting $t = r$.
		\subsection{Mean Popularity of Predicted Cascade Size}
		To determine the mean tree size of such a discussion, we can differentiate Eq.~\eqref{Ipgf} and evaluate at $x = 1$ as in Sec.~\ref{sec:model}. This gives
		\begin{align}
		m_I(r) &= \left.\pd{I\left(\{a_0,a_1,\dots,a_{n-1}\}, r; x\right)}{x}\right|_{x=1} \\
		&= n + \int_{0}^{r}\left(\mu(r+a_0-w) + \sum_{i=1}^{n-1} \xi \, \p(r+a_i-w)\right) \, m_G(w)\, d w,
		\label{Imean}
		\end{align}
		where the $n$ term represents the events that occurred during the observation window and $m_G(w)$ represents the mean tree size of prediction age $r$ and is given by Eq.~\eqref{mean_g}. Observing that
		\begin{equation}
		\xi \, \int_{0}^{r} \phi(r + a_i - w)\,m_G(w)\,d w
		\label{phi_contr}
		\end{equation}
		can be written as a convolution integral in terms of the quantity $\Psi_i(r)$ defined by $\Psi_i(r) = \phi(r+a_i)$, we note that the term in Eq.~\eqref{phi_contr} has Laplace transform  $\xi\, \hat{\Psi_i}(s) \, \hat{m}_G(s)$. Applying this to each term in the summation of Eq.~\eqref{Imean} gives the Laplace transform of $m_I(r)$ as
		\begin{equation}
		\hat{m}_I(s) = \frac{n}{s} + \frac{1}{s}\left[\frac{\hat{\Phi}(s) + \sum_{i=1}^{n-1} \xi \, \hat{\Psi}_i(s)}{1 - \xi\,\hat{\phi}(s)}\right],
		\label{meanIlap}
		\end{equation}
		where
		\begin{align}
		\hat{\Psi}_i(s) &= \int_{0}^{\infty}\phi_i(r + a_i) \, e^{-rs} \, d r, \\
		&= e^{a_is}\left[\hat{\phi}(s) - \int_{0}^{a_i}\phi(\tilde{r}) \, e^{-\tilde{r}s} \, d \tilde{r}\right],
		\label{Lpsi}
		\end{align}
		using the change of variable $\tilde{r} = r + a_i$. Similarly, we define $\Phi(r) = \mu(r + a_0)$ such that
		\begin{equation}
		\hat{\Phi}(s) = e^{a_0 s}\left[\hat{\mu}(s) - \int_{0}^{a_0}\mu(\tilde{r}) \, e^{-\tilde{r}s} \, \dd \tilde{r}\right].
		\label{LPhi}
		\end{equation} 
		The evaluation of this transform now depends only on (a finite) numerical integration, provided we know the Laplace transform of both $\phi$ and $\mu$. Thus, this approach is extremely feasible. Furthermore, we can also determine some information regarding the large-$r$ behavior of the mean via the final-value theorem to obtain
		\begin{align}
		\lim\limits_{r\to \infty}m_I(r) &= \lim\limits_{s\to 0} \, s \, \hat{m}_I(s) \\
		&= n + \left(\frac{\hat{\Phi}(0) + \xi \, \sum_{i=1}^{n-1}\hat{\Psi}_i(0)}{1 - \xi}\right),
		\end{align}
		where by Eq.~\eqref{Lpsi}
		\begin{equation}
		\hat{\Psi}_i(0) = 1 - \int_{0}^{a_i} \phi(\tilde{r}) \, d{\tilde{r}},
		\end{equation}
		and similarly
		\begin{equation}
		\hat{\Phi}(0) = 1 - \int_{0}^{a_0}\mu(\tilde{r}) \, d\tilde{r},
		\end{equation}
		which allows us to obtain
		\begin{equation}
		\lim_{r \rightarrow \infty} m_I(r) = n + \left(\frac{1 - \int_{0}^{a_0}\mu(\tilde{r}) \, d \tilde{r} + \xi \, \sum_{i=1}^{n-1} \left[1 - \int_{0}^{a_i} \phi(\tilde{r}) \, d{\tilde{r}}\right]}{1-\xi}\right).
		\end{equation}
		
		\subsubsection{Exponential memory with constant background intensity}
		
		We now consider, as in Sec.~\ref{sec:mean}, the specific case of an exponential memory function along with some constant background intensity [i.e., $\mu(t) = \lambda_0$].
		In this case $\phi(t) = \beta e^{-\beta t}$ and therefore $\hat{\phi}(s) = \frac{\beta}{\beta + s}$. Equation (\ref{Lpsi}) in this scenario is given by
		\begin{align}
		\hat{\Psi}_i(s) &= e^{a_is}\left[\frac{\beta}{\beta + s} - \beta \int_{0}^{a_i}e^{-\beta \tilde{r}}e^{-\tilde{r}s} \, d \tilde{r}\right] \\
		&= \frac{\beta e^{-\beta a_i}}{\beta + s},
		\end{align}
		and therefore
		\begin{equation}	
		\hat{m}_I(s) = \frac{n}{s} + \frac{1}{s\left[\beta(1-\xi) + s\right]}\left[\frac{\lambda_0(\beta + s)}{s} + \xi \, \sum_{i=1}^{n-1} \beta e^{-\beta a_i}\right].
		\label{Lap_mi}
		\end{equation}
		This allows us to write the following closed-form expression for the expected cascade size at observation time $r$:
		\begin{equation}
		m_I(r) =	n +  \frac{\lambda_0}{1 - \xi}\left[r + \frac{\xi}{\beta(1-\xi)}\left(e^{-\beta(1-\xi)r} - 1\right)\right] +  \sum_{i=1}^{n-1}  \frac{\xi \, e^{-\beta a_i}}{1-\xi}\left[1 - e^{-\beta(1-\xi)r}\right].
		\label{mean_pre_exp}
		\end{equation}
		Note that this equation consists of three components: (i) the number of events we have observed at the start of the prediction window $n$; (ii) the contribution from the background intensity as we saw in Eq.~\eqref{mean_exp_classic}, reflecting the constant background intensity; and (iii), the contribution from the events in the cascade during the observation period. We would like to highlight once more the fact that in the presence of $\lambda_0 \ne 0$, the expected number of events increases linearly with time with a non zero probability of events occurring at each time step. In contrast, we next consider a decaying background intensity.
		
		\subsubsection{Exponential background intensity and memory kernel} 
		Suppose that the background intensity is decaying with time such that $\mu(t) = \alpha e^{-\alpha t}$, while the memory kernel is the same as in the previous case, i.e., exponentially distributed with mean time $1/\beta$. Equation \eqref{LPhi} is then given by
		\begin{align}
		\hat{\Phi}(s) &= \frac{\alpha e^{-\alpha(a_0)}}{\alpha + s},
		\end{align}
		where $a_0 = T$ is the age of the seed comment when the observation period ends. We can obtain the expected number of events at time $r$ in the prediction interval by inverting Eq.~\eqref{meanIlap},
		\begin{align}
		m_I(r) = n &+ \alpha e^{-\alpha a_0}\left\{\frac{1}{\alpha(1-\xi)} + e^{-\alpha r}\left[\frac{\alpha - \beta}{\alpha\left[(1-\xi)\beta - \alpha\right]}\right] \right. \nonumber \\ &- \left.e^{-\alpha(1-\xi)r}\left[\frac{\xi}{(1-\xi)(\alpha - [1 - \xi]\beta)}\right]\right\} + \sum_{i=1}^{n-1}  \frac{\xi \, e^{-\beta a_i}}{1-\xi}\left[1 - e^{-\beta(1-\xi)r}\right].
		\end{align}
		Note that, unlike previous examples such as Eq.~\eqref{mean_pre_exp}, this quantity has a finite limit as $r \rightarrow \infty$; this is to be expected as there is no longer a constant probability of an event occurring in each time step but rather a decaying probability and as such the likelihood of events occurring after long inter event times is significantly reduced. Thus, we may calculate the large-$r$ expected number of events to be 
		\begin{equation}
		m_I(\infty) = n + \frac{1}{1-\xi}\left[e^{-\alpha a_0} + \sum_{i=1}^{n-1} e^{-\beta a_i} \right].  
		\end{equation}
		\subsection{Probability a discussion receives no further comments}
		In a similar fashion as in Sec.~\ref{subsubsec:noComments}, we can calculate the probability that the discussion tree receives no further comments in the prediction window by time $r$, $q_n(r)$, as follows:
		\begin{equation}
		q_n(r) = \lim\limits_{x \to 0}\frac{I\left(\{a_0,a_1,\dots,a_{n-1}\}, r; x\right)}{x^n}\,.
		\end{equation}
		Since $q_m(r) = 0, \,\, \forall m < n$, this results in
		\begin{equation}
		q_n(r) = \exp\left\{-\left[\int_{0}^{r} \mu(r+a_0-w) + \xi \sum_{i=1}^{n-1} \phi(r+a_i-w) \, d w\right]\right\}.
		\end{equation}
		We can again calculate this quantity exactly in the case of exponential memory with constant background intensity to obtain
		\begin{equation}
		q_n(r) = \exp\left\{-\left[\lambda_0 r + \xi \sum_{i=1}^{n-1} e^{-\beta a_i}\left(1 + e^{-\beta r}\right)\right]\right\}\,.
		\label{q1_exp_pred}
		\end{equation}
		Note that whereas in Eq.~\eqref{q1classic} the probability of receiving no more comments depended only on exogenous events, in this expression we explicitly account for the endogenous events in the history of the process prior to time $T$.
		
		\section{Comparison of Numerical Simulations with Theory}\label{sec:simulations}
		
		In order to validate these expressions, as well as to test their predictive power, we now proceed to compare them with numerical simulations of Hawkes processes. In each simulation we observe all events that took place before a certain time $T$, i.e., at $\{\tau_0,\tau_1,\ldots,\tau_{n-1}\}$ with $\tau_{n-1}<T$. Then we predict the evolution of the discussion tree until time $\Omega$, using the PGF describing the distribution of future events from Eq.~\eqref{Ipgf}, which we again invert via fast Fourier transforms as in Sec.~\ref{num_sim_1}.
		
		Figure \ref{fig:sim_errorbars} shows the case of a Hawkes processes with constant background intensity and an exponential memory kernel [Fig.~\ref{fig:sim_errorbars}(a)] or a power-law memory kernel [Fig.~\ref{fig:sim_errorbars}(b)]. In each case, we observe a time series of events until time $T = 10$. Then we make a prediction of the dynamics until time $\Omega = 20$. Specifically, in each panel, we calculate the ensemble average as well as the 95th percentile range, at each time step over all ensembles. This is compared with the theoretical expected number of events as well as the 95\% prediction interval with the information available up to time $T$ obtained via numeric inversion of Eq.~\eqref{Ipgf} via the fast Fourier transform method described in Sec.~\ref{sec:IFFT}.
		
		\begin{figure}[t!]
			\centering
			\includegraphics[width=\linewidth]{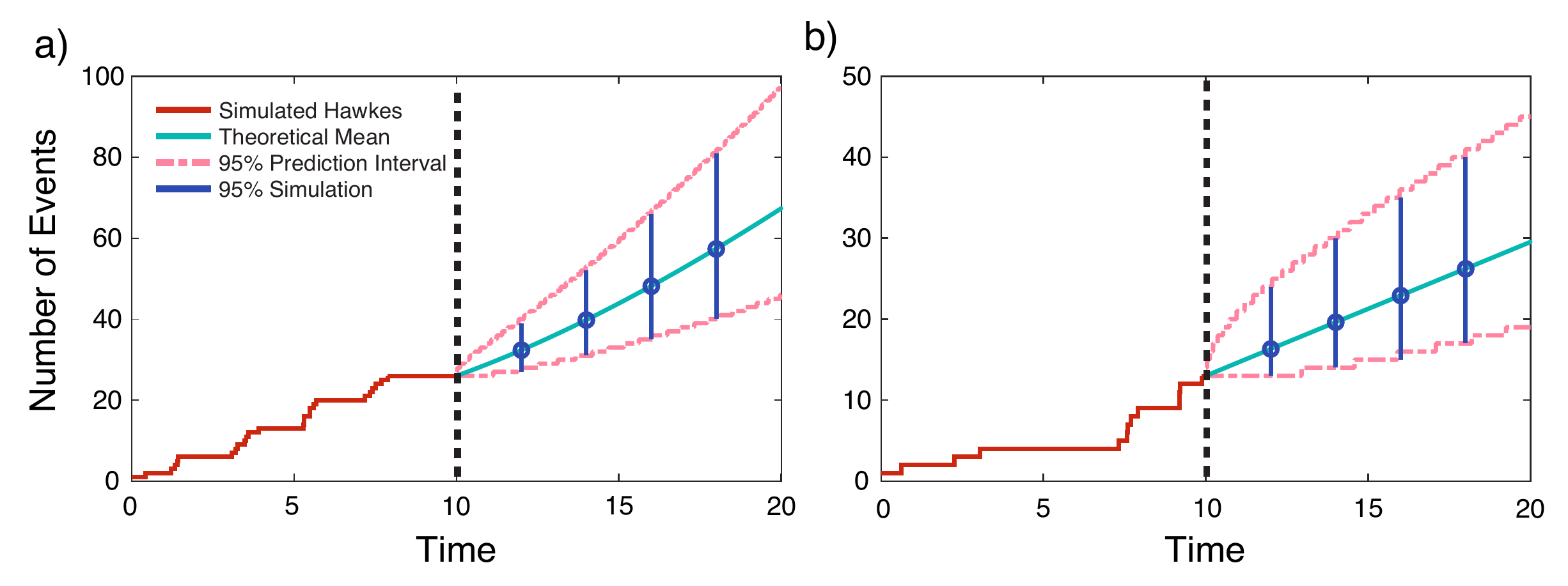}
			\caption{Numerical simulation and theoretical results for two Hawkes processes where one realization is obtained over the interval $[0,10]$ before proceeding to simulate $10^6$ ensembles based upon these events over the interval $(10,20]$. We show the average number of events along with the 95\% percentiles from the distribution of events at a number of time points. In addition, we show the theoretical expected number of events and 95\% prediction intervals obtained from numerical inversion of Eq.(\ref{Ipgf}). (a) Case of background intensity $\lambda_0 = 1.5$, fitness $\xi = 0.8$, and exponential memory kernel with mean time 3 ($\beta=1/3$). (b) Case with shifted power-law memory kernel ($\beta = 1, c = 0.01, \xi = 0.5$, and $\lambda_0 = 1$).}
			\label{fig:sim_errorbars}
		\end{figure}
		
		The first thing we observe is that, during the observation window, in the exponential case the frequency of events is much more stable over time in comparison to the bursty nature of the power-law memory kernel, which is characterized by larger periods of inactivity followed by a number of events over a short time interval  \cite{barabasi2005origin, karsai2018bursty}. We also point out the difference between the prediction intervals between the two distributions. Nevertheless, the agreement in both cases between the theoretical predictions and the numerical simulations is exceptionally good.
		
		To determine how the age of comments at time $T$ affects the future popularity of the post, we generate two synthetic time series of events for the observation period each containing ten events. Then we simulate the rest of the process with those series acting as the history of the content and predict the evolution using our analytical tools with results shown in Fig.~\ref{fig:diff_times}. In Fig.~\ref{fig:diff_times}(a) the synthetic series consists of equally spaced inter event times at $\{0,1,2,3,4,5,6,7,8,9\}$, while the second sequence shown in Fig.~\ref{fig:diff_times}(b) is such that there are also ten events but some comments are ``younger'' at time $T$, $\{0,1,2,3,5,8,9,9.5,9.6,9.8\}$. Between times $T$ and $\Omega = 20$ we then proceed to simulate the Hawkes process given by Eq.~\ref{expHawkes} with parameters $\beta = 1/3$, $\lambda_0 = 0.1$, and $\xi = 0.8$. In both cases, we show the theoretical mean given by Eq.~\eqref{mean_pre_exp}, the 95\% prediction intervals drawn from the distribution at each time step, and also the mean measured directly from the distribution obtained numerically rather than the analytical equation. As we can see, the occurrence of younger posts when making predictions of future popularity results in larger predicted cascade sizes, correctly matched with the results obtained from the simulations both for the expected value and for the upper limits of the prediction intervals.
		
		Also shown in Figs.~\ref{fig:diff_times}(c) and \ref{fig:diff_times}(d) are the corresponding CCDFs, which are the probability of having $m$ or more comments on the discussion tree,  at different times. Again, as expected, the larger the age of a given tree, the more likely it is to have received more posts. Also, the process with younger comments by the end of the observation time has higher probability of having more popular discussions due to the time-decaying memory kernel.
		
		\begin{figure}[t!]
			\centering
			\includegraphics[width=\linewidth]{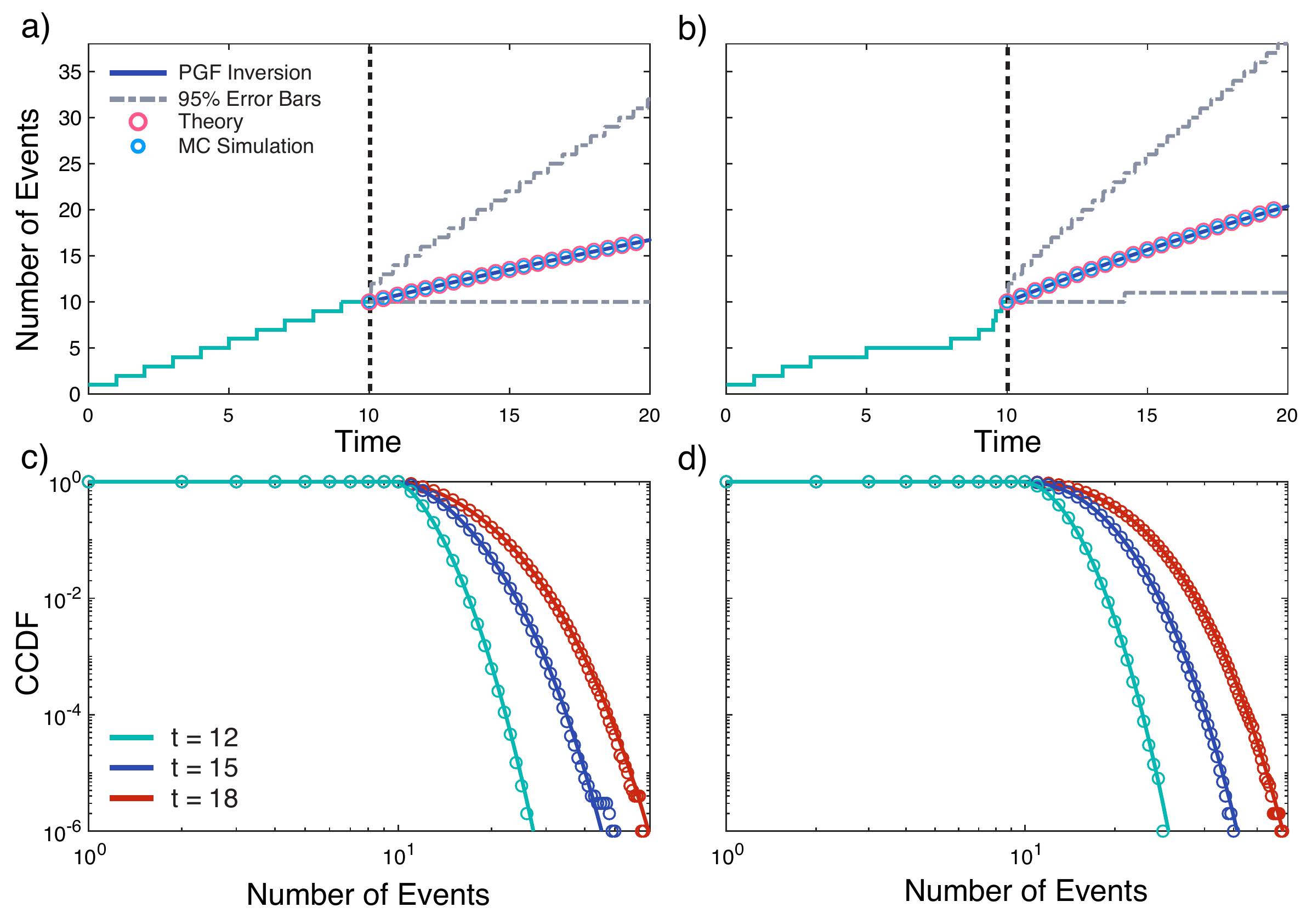}
			\caption{Comparison of numerical simulations of two half-synthetic Hawkes processes with analytical results. In both cases, the constant background activity is $\lambda_0=0.1$ and the excitation function is given by an exponential distribution with mean time 3 ($\beta=1/3$) and fitness $\xi=0.8$. Moreover, in both cases, we observe $n=10$ events before $T$, although in this case these events are generated synthetically so that (a) $\tau=\{0,1,2,3,4,5,6,7,8,9\}$ and (b) $\tau=\{0,1,2,3,5,8,9,9.5,9.6,9.8\}$. Then the rest of the process is numerically computed and the prediction of future events is done analytically up to time $\Omega=20$. The mean number of events is shown in (a) and (b) for the different observation times. In (a) and (b) the circles represent the average over $10^6$ Monte Carlo simulations, while the blue line shows the theoretical mean calculated directly from the inversion of the PGF given by Eq.~\eqref{Lap_mi}. Note that for this particular case of a Hawkes process we may also obtain an analytical expression for this average through Eq.~\eqref{mean_pre_exp}. We also show the $95\%$ prediction intervals for the popularities at each time point with the gray lines, again via inversion of Eq.~\eqref{Lap_mi}. (c) and (d) The CCDFs for the number of events at a number of different time points, where the lines are the theoretical distributions and the circles are observed CCDFs from the numerical simulations.}
			\label{fig:diff_times}
		\end{figure}
		
		Finally, in Fig.~\ref{fig:q1_diff_eps} we again consider the evenly spaced ten events prior to time $T = 10$ as per Fig.~\ref{fig:diff_times}(a). In Fig.~\ref{fig:q1_diff_eps}(a) we explore the effect of different fitness parameters ($ \xi = 0.2,0.5,0.8$) for the discussion's future popularity, while in Fig.~\ref{fig:q1_diff_eps}(b) we show the probability $q_n(r)$, for each of the above fitness values, that the tree does not receive another comment between $T$ and $\Omega$, as given by Eq.~\eqref{q1_exp_pred}. The agreement between theory and numerical simulations is again excellent.
		
		\begin{figure}[t!]
			\centering
			\includegraphics[width=\linewidth]{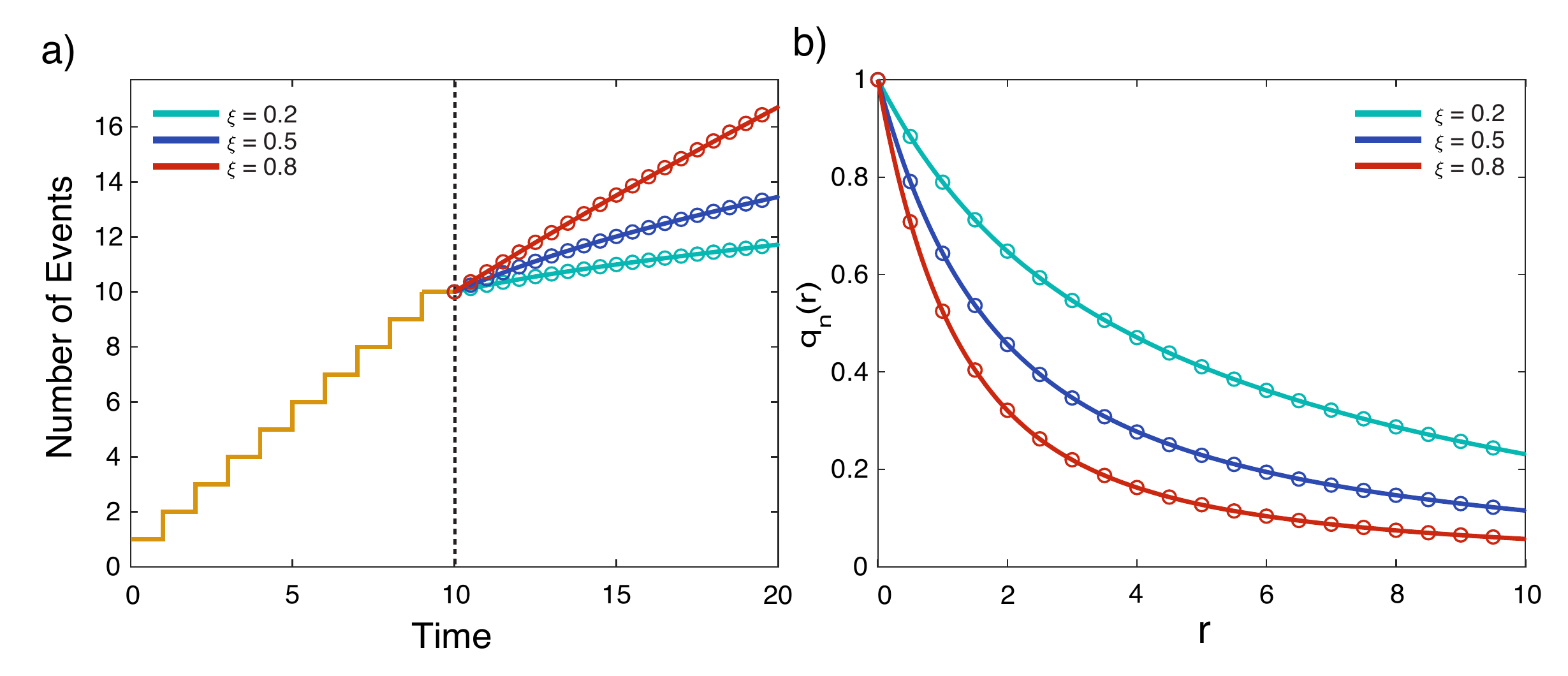}
			\caption{Simulations equivalent to the one in Fig.~\ref{fig:diff_times}(a), i.e., with a synthetic evenly spaced series of events before $T=10$, but with different fitness values for the rest of the evolution up to time $\Omega$. (a) Observed series along with the mean number of events during the prediction window. (b) Probability of zero replies (posts) in a prediction window of length $r$, $q_n(r)$.}
			\label{fig:q1_diff_eps}
		\end{figure}

		\newpage
		
		\section[Conclusions]{Conclusions}\label{sec:conclusions}
		In this paper we have introduced an analytically tractable predictive model built upon a branching process interpretation of the most general form of the Hawkes process, namely, where both the background intensity and the memory kernel are arbitrary time-dependent functions. The focus on the underlying branching process model is important for the generality of the method: While we have considered the dynamics of online opinion boards when describing this model, the bounds on predictability given by the theory are applicable to a wide range of phenomena. Examples of phenomena that can be described by such processes include (but are not limited to) natural catastrophes such as earthquakes \cite{ogata1988statistical,ogata1998space,saichev2006universal,ogata2002modern,ogata1988statistical, helmstetter2002subcritical}, the popularity of Youtube videos \cite{crane2008robust}, criminal behavior \cite{mohler2011self}, and conversation frequency \cite{masuda2013self}. In comparison to related works, a crucial feature of our analysis is that not only does it give the expected number of future events but it also (in a mathematically consistent way) bounds the predictability in the sense of deriving the entire probability distribution of future events. In applications, this allows the direct calculation of not only the expected number of events but also the entire theoretical distribution, without the need for extensive Monte Carlo simulations to determine the time evolution of the process. In turn, this opens the path to the development of, possibly unsupervised, methods that could evaluate and predict expected values (and their confidence intervals) for quantities associated with self-exciting dynamics such as cascade sizes. 
		
		The strength of this model's predictions is, like that of all models, limited by the accuracy of the parameters estimated from data to describe the process. Issues regarding the estimation of such parameters to describe empirical data are well-known and have been highlighted along with potential remedies in numerous pieces of literature, for example \cite{rizoiu2017hawkes, rasmussen2013bayesian, daley2003introduction}. In order to apply the theory introduced in this article to a wide range of applications, we suggest that, while beyond the scope of the present article, future work should focus on improved approaches in estimating parameters, which importantly may in fact be time dependent, as suggested in this article, for both the background intensity and/or the fitness parameter of the memory kernel.
		
		In summary, we believe that the theoretical approach developed in this paper creates an opportunity for more careful prediction of cascading dynamics in online social media. The ability to predict the entire distribution of possible future popularity values means that confidence limits can be placed on predictions, which should allow practitioners to better appreciate the underlying stochasticity of human behavior and to gain some measure of control over its influence on dynamics. Future work will also be devoted to a better characterization of real processes, such as those related to posting on opinion boards, with the aim of learning which kind of background intensity and kernel are best suited to described these processes accurately and how they are correlated with human (bursty) activity.

		
		\begin{acknowledgements}
			We thank Kevin Burke, Renaud Lambiotte, and David O'Sullivan for helpful discussions and comments. This work was supported by Science Foundation Ireland Grant No. 16/IA/4470, No. 16/RC/3918, No. 12/RC/2289P2 and No. 18/CRT/6049 (J.D.O'B. and J.P.G.). J.D.O'B. acknowledges support from the Mathematics for Industry Network (Grant No. ECOST/STSM/TD1409/290216/071429). A.A. acknowledges the support of the FPI doctoral fellowship from MINECO (Grant No. FIS2014-55867-P). A.A. and Y.M. acknowledge partial support from the Government of Aragon, Spain through Grant No. E36-17R (FENOL), and from MINECO and FEDER funds (Grant No. FIS2017-87519-P). A.A and Y.M also acknowledges support from Intesa Sanpaolo Innovation Center. We acknowledge the DJEI/DES/SFI/HEA Irish Centre for High-End Computing for the provision of computational facilities and support. 
		\end{acknowledgements}	
		
		\appendix
		\numberwithin{equation}{section}
		\section{Derivation of the Bellman-Harris Integral Equation}\label{app:BH_derivation}
		In this appendix we show how the integral equation of the well-known Bellman-Harris Process \cite{Athreya2004,Harris2002,Iribarren2011_PRE} may be derived via our differential equation branching process approach. As in the classical discrete-time Galton-Watson model, a single particle is born at time $t = 0$ but rather than living for a single time unit, the particles live until an age $a$, which is a random variable with distribution $\gamma(a)$. From this definition we may also determine $\lambda(a)$ such that $\lambda(a) \Delta t$ describes the probability that a particle alive at age $a$ dies in the age interval $(a,a+\Delta t)$, or the so-called hazard rate. At the moment of death, the particle produces $k$ offspring or progeny with probability $p_k$ each of which behaves exactly as their parent i.e., each particle is independent and identically distributed. In the following, we will use the random variable $k$ through its PGF
		\begin{equation}
		f(x) = \sum_{k = 0}^{\infty} p_k x^k.
		\end{equation}
		Due to the previous considerations, these processes are known as age-dependent continuous-time branching processes.
		
		We proceed, as in the main text, to define $G(\tau,a,\Omega;x)$ to be the PGF for the distribution of the number of particles, observed at time $\Omega$, that have lived as a result of a parent of age $a$ producing offspring at time $\tau$. Note that we previously described this as the size of a tree. However, in the specific case of the Bellman-Harris process, every particle dies at the time of seeding a new tree. Now consider how this PGF may change over infinitesimal time interval $(\tau-\Delta t, \tau)$, where $\Delta t$ is sufficiently small to ensure only one event at most may occur in this interval. There are two possibilities.
		\begin{enumerate}[(i)]
			\item The parent may die with probability $\lambda(a) \Delta t$ producing a random number $k$ children, each of which may start their own subtree at time $\tau$ but with age $a = 0$. As we are concerned with the total number of particles to have lived in this interval we must also count the (now deceased) parent particle through an additional power of $x$ in the PGF and as such the total contribution to $G(\tau-\Delta t,a - \Delta t, \Omega; x)$ is $x f\left[G(\tau,0,\Omega;x)\right]$ .
			\item The parent may also live, which occurs with probability $1 - \lambda(a) \Delta t$, and the contribution to $G(\tau - \Delta t,a - \Delta t,\Omega;x)$ in this case is simply $G(\tau,a,\Omega;x)$. 
		\end{enumerate}
		Based on the above, the equation governing the change in $G$ over the time interval $(\tau - \Delta t, \tau)$ is
		\begin{equation}
		G(\tau-\Delta t,a - \Delta t, \Omega; x) = \lambda(a) \, x f\left[G(\tau,0,\Omega;x)\right] \Delta t + \left[1 - \lambda(a) \Delta t\right] \, G(\tau,a,\Omega;x),
		\end{equation}
		which when taking the $\Delta t \rightarrow 0$ limit and writing $G(\tau,a,\Omega;x)$ as $G$ becomes
		\begin{equation}
		\pd{G}{\tau} + \pd{G}{a} = \lambda(a) \left\{G - x f\left[G(\tau,0,\Omega;x)\right]\right\}.
		\end{equation}
		Now, as in the main text, we use the method of characteristics with $c = \tau - a$ (being a constant along a characteristic) to express the change in $G$ along a characteristic as
		\begin{equation}
		\frac{d G}{d \tau} = \lambda(\tau - c) \left\{G - x f\left[G(\tau,0,\Omega;x)\right]\right\},
		\end{equation}
		which, along with the final condition $G(\Omega,a,\Omega;x) = x$, i.e., a tree seeded at observation time has size one, may be solved to give
		\begin{align}
		G(\tau,a,\Omega;x) &= x \, \exp\left[- \int_{\tau}^{\Omega}\lambda(w - \tau + a) \, d w\right] \nonumber
		\\ 
		& + \qquad x \, \int_{\tau}^{\Omega}\exp\left[-\int_{\tau}^{t'}\lambda(w - \tau + a) \, d w\right] \lambda(t' - \tau + a) f\left[G(t',0,\Omega;x)\right] d t'.
		\end{align}
		Introducing the tree-age $t = \Omega - \tau$, i.e., the time between a tree being seeded and observed, and expressing $G(\tau,a,\Omega,x)$ as $G(\Omega - \tau, a; x)$, the above equation becomes
		\begin{align}
		G(t,a;x) &= x \, \exp\left[- \int_{0}^{t}\lambda(y + a) \, d y\right] \nonumber \\
		& + \qquad x \, \int_{0}^{t}\exp\left[-\int_{0}^{s}\lambda(u + a) \, d u\right] \lambda(s + a) f\left[G(t-s, 0 ;x)\right] d s,
		\end{align}
		where $y = w - \tau$ and $s = t' - \tau$. The above equation describes the PGF for the size of a tree with tree-age $t$ seeded by a parent of age $a$. Finally, as we are concerned with the total number of particles in a given process which starts through a single parent being born we set $a = 0$ (and dropping the $a=0$ argument from $G$) to obtain
		\begin{equation}
		G(t;x) = x \, \exp\left[- \int_{0}^{t}\lambda(y) \, d y\right] + x \, \int_{0}^{t}\exp\left[-\int_{0}^{s}\lambda(u) \, d u\right] \lambda(s) f\left[G(t-s;x)\right] d s,
		\end{equation}
		which we may express as
		\begin{equation}
		G(t;x) = x \, S(t) + x \int_{0}^{t} \gamma(s) f\left[G(t-s;x)\right] d s,
		\label{BH}
		\end{equation}
		where $\gamma(a)$ describes the probability of living until age $a$ while $S(a) = \int_{a}^{\infty} \gamma(s) \, d s$ gives the probability of a particle having a lifetime of length $a$ or longer, which is the classical Bellman-Harris integral equation.
		
		On a final note, we mention that from Eq.~\eqref{BH} we may derive the governing equations for two more specific branching processes. First, we obtain the Markovian age-dependent branching process equation through letting $\lambda(a) = \lambda$, i.e., a constant probability of death in each time interval. Second, we can describe the classical Galton-Watson branching process by ensuring each particle has a lifetime of $1$, i.e., $\gamma(a) = \delta_{a,1}$, a Dirac delta function with mass at $a = 1$.
		
		\bibliographystyle{unsrt}
		\bibliography{hawkes_bib}
		
		
	\end{document}